\documentclass{article}
\usepackage{spconf,amsmath,graphicx}

\usepackage{subfigure} 
\graphicspath{{./image/}}

\usepackage{amssymb}

\DeclareMathOperator*{\softmax}{softmax}
\DeclareMathOperator*{\sigmoid}{sigmoid}
\usepackage{epstopdf}
\usepackage{tabularx}
\usepackage{multirow}
\usepackage{textcomp}
\usepackage{url}

\usepackage{bm}
\usepackage{calc}

\usepackage{balance}

\usepackage{color}

\title{ENABLING EARLY AUDIO EVENT DETECTION WITH NEURAL NETWORKS}
 \name{Huy Phan\sthanks{The work was performed when H. Phan was at the Institute of Signal Processing, University of L\"ubeck.},
	Philipp Koch$^{\ddagger}$, 
	Ian McLoughlin$^{\dagger}$, and
	Alfred Mertins$^{\ddagger}$}
\address{$^\ast$ University of Oxford, Department of Engineering Science, Oxford, UK \\
	$^\dagger$ University of Kent, School of Computing, Kent, UK \\
	$^\ddagger$ University of L\"ubeck, Institute for Signal Processing, L\"ubeck, Germany \\
	{\small \tt huy.phan@eng.ox.ac.uk, ivm@kent.ac.uk, {koch,mertins}@isip.uni-luebeck.de}}

\begin{document}
\ninept
\maketitle
\begin{abstract}
This paper presents a methodology for early detection of audio events from audio streams. Early detection is the ability to infer an ongoing event during its initial stage. The proposed system consists of a novel inference step coupled with dual parallel tailored-loss deep neural networks (DNNs). The DNNs share a similar architecture except for their loss functions, i.e. weighted loss and multitask loss,  which are designed to efficiently cope with  issues common to audio event detection.
%The weighted-loss DNN is used to determine whether an input audio frame is foreground or background. The multitask-loss DNN, which is a joint classification-regression model, is employed for target event classification. 
The inference step is newly introduced to make use of the network outputs for recognizing ongoing events. The monotonicity of the detection function is required for reliable early detection, and will also be proved. 
Experiments on the ITC-Irst database show that the proposed system achieves state-of-the-art detection performance. Furthermore, even partial events are sufficient to achieve good performance similar to that obtained when an entire event is observed, enabling early event detection.
\end{abstract}
\vspace{-0.2cm}
\section{Introduction}
\label{sec:intro}
\vspace{-0.3cm}

Great progress has been made in recent years on the problem of audio event detection, in both methodologies \cite{McLoughlin2015,McLoughlin2017,Cakir2017,Takahashi2016a,phan2016c} as well as available datasets \cite{dcase2016web,dcase2017web,Mesaros2017,Mesaros2016}. 
However, many previous works focused only on classifying audio events after fully observing an entire event. 
We are still missing an important aspect of audio event detection: namely, the early detection of ongoing events, such as from live audio streams. 
Early AED differs from standard AED by requiring ongoing events to be recognized as early as possible, or when given only a partial observation of the beginning of an event~\cite{phan2015b}. 
The ability to \emph{reliably} detect ongoing events at their early stage is important in many scenarios, such as surveillance and safety-related applications, which require low latency reaction to potentially dangerous events. % as early as possible or to forecast dangerous situation. 
However, this requires monotonicity in the detection function, which is not easily fulfilled with methods used to recognize complete events~\cite{phan2015b,Hoai2014}. 

\begin{figure} [!t]
	\centering
	\includegraphics[width=0.95\linewidth]{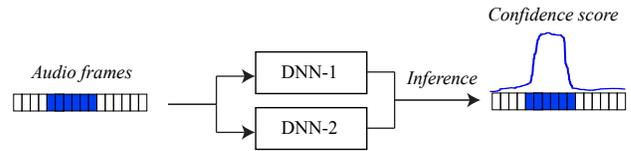}
	\vspace{-0.3cm}
	\caption{An overview of the proposed system.}
	\label{fig:pipeline}
	\vspace{-0.4cm}
\end{figure}

There has been an influx of works employing deep networks, such as DNNs \cite{McLoughlin2015,Laffitte2016}, convolutional neural networks (CNNs) \cite{McLoughlin2017,Kumar2017,Takahashi2016a,Zhang2015}, and recurrent neural networks (RNNs) \cite{Cakir2017,Parascandolo2016}, for audio event analysis. 
However, these works primarily focused on new network architectures. Little attention has been paid on loss functions to address the common issues of audio event detection in audio streams and, more importantly, to enable inferring an ongoing event at its early stage. 
To this end, we propose two tailored loss functions: (1) weighted loss for unbalanced foreground/background classification and (2) multitask loss for jointly modelling event class distributions (i.e. event classification) and event temporal structures (i.e. regression for event boundary estimation). 
These loss functions are used with two DNNs which are operating in parallel as demonstrated in Fig.~\ref{fig:pipeline}. Furthermore, a novel inference scheme is introduced to make use of the network outputs to infer ongoing events for early detection. Intuitively, the confidence score of a given target event occurring %at a time index 
will gradually accumulate when each incoming audio frames describing that event is analysed. 
Since the confidence score is a monotonic function (see Section \ref{ssec:monotonicity}), an event that is detectable by the system can be reliably detected early in time as soon as the accumulated confidence score reaches a predetermined threshold. 
%Introduction goes here ... 
%Discuss relation to other neural networks \cite{McLoughlin2015,McLoughlin2017,Zhang2016}

%Our work builds upon our previous deep learning innovations~\cite{Phan2015, phan2015b, Phan2016e}. 
This work can be seen as an extension of our previous work \cite{Phan2015, phan2015b, Phan2016e} into the deep learning paradigm.
%Our work builds upon our previous deep learning innovations~\cite{Phan2015, phan2015b, Phan2016e}. 
In \cite{Phan2015, phan2015b}, different decision forests were learned separately for event classification and event boundary estimation. 
Although these two tasks can be jointly learned with classification-regression forests \cite{Phan2016e}, they are class-specific. In contrast, the proposed multitask DNN is a single model for multiple classes and multiple tasks simultaneously. It is also worth mentioning that the tailored loss functions were used by our AED system \cite{Phan2017e} for the DCASE 2017 challenge \cite{dcase2017web}. However, the simple median filtering used in that preliminary work is not capable of early event detection. The novel inference scheme proposed in this work is explicitly designed to overcome this limitation.

\vspace{-0.2cm}
\section{The Proposed Approach}
\label{sec:proposed_approach}
\vspace{-0.3cm}

\begin{figure} [!t]
	\centering
	\includegraphics[width=0.75\linewidth]{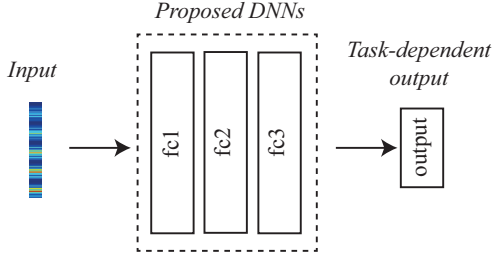}
	\vspace{-0.2cm}
	\caption{The proposed DNN architecture.}
	\label{fig:dnn}
	\vspace{-0.25cm}
\end{figure}

An overview of the proposed system is illustrated in Fig. \ref{fig:pipeline}. A continuous audio signal is firstly decomposed into frames. Each frame is then presented to the weighted-loss and multitask-loss DNNs denoted as \emph{DNN-1} and \emph{DNN-2}, respectively. 
The former is used to determine whether the input frame is foreground or background whereas the latter is employed for joint event classification and event boundary estimation. 
DNN-1 and DNN-2 share a similar architecture, including three fully connected layers as demonstrated in Fig. \ref{fig:dnn} and Table \ref{tab:dnn}. 
The differences are the dropout probability \cite{Srivastava2014}, which is 0.5 and 0.2 for DNN-1 and DNN-2 respectively, the output layer, and the loss function. Note that the DNNs operate in parallel rather than as a cascade in \cite{Phan2017e,phan2015b,Phan2015}. The network outputs are then used in the inference step to compute a confidence score that a target event is occurring at a certain time index.

\vspace{-0.3cm}
\subsection{DNN-1: Fore-/background classification with weighted loss}
\label{ssec:background_rejection}
\vspace{-0.1cm}

In general, for audio event detection in continuous streams, the number of background frames is significantly larger than of foreground ones. This leads to a skewed classification problem with a dominance of the background samples. Since foreground samples are more valuable than background ones, we penalize the network more for false negative errors than for false positives. The weighted loss is designed for this purpose.

Let $\left\{\left(\mathbf{x}_1,\mathbf{y}_1\right),\ldots,\left(\mathbf{x}_N,\mathbf{y}_N\right)\right\}$ denote a training set of $N$  examples where $\mathbf{x}\in \mathbb{R}^{D}$ denotes a feature vector of size $D$. $\mathbf{y} \in \{0,1\}^{2}$ denotes a binary one-hot encoding vector where $0$ and $1$ indicate background and foreground categories, respectively. The weighted loss reads,
\vspace{-0.2cm}
\begin{align}
	E_{\mbox{\scriptsize w}}(\boldsymbol{\theta}) &= -\frac{1}{N}\bigg(\lambda_{\mbox{\scriptsize fg}}\sum_{n=1}^{N}{\mathbb{I}_{\mbox{\scriptsize fg}}(\mathbf{x}_n)\mathbf{y}_n\log\big(\mathbf{\hat{y}}_n(\mathbf{x}_n\,,\boldsymbol{\theta})\big)} \nonumber \\
	& + \lambda_{\mbox{\scriptsize bg}}\sum_{n=1}^{N}\mathbb{I}_{\mbox{\scriptsize bg}}(\mathbf{x}_n){\mathbf{y}_n\log\big(\mathbf{\hat{y}}_n(\mathbf{x}_n\,,\boldsymbol{\theta})\big)}\bigg) + \frac{\lambda}{2}\left\|\boldsymbol{\theta}\right\|^2_2,
	\label{eq:weighted_loss}
	\vspace{-0.6cm}
\end{align}
where $\boldsymbol{\theta}$ denotes the network's trainable parameters. $\mathbb{I}_{\mbox{\scriptsize fg}}(\mathbf{x})$ and $\mathbb{I}_{\mbox{\scriptsize bg}}(\mathbf{x})$ are indicator functions which specify whether the sample $\mathbf{x}$ is foreground or background, respectively. $\lambda_{\mbox{\scriptsize fg}}$ and $\lambda_{\mbox{\scriptsize bg}}$ are \emph{penalization weights} for false negative errors 
%(i.e. a foreground sample is misclassified as background) 
and false positive errors 
%(i.e. a background sample is misclassified as foreground)
, respectively. The hyper-parameter $\lambda$ is used to trade off the error terms and the {$\ell_2$-norm} regularization term, $\left\|\boldsymbol{.}\right\|^2_2$. The  posterior probability $\mathbf{\hat{y}}(\mathbf{x}\,,\boldsymbol{\theta})$ is obtained by applying a softmax to the  output layer.

%-----------------------------------
\begin{table}[b!]
	\caption{The parameters of the DNNs. A dropout probability of 0.5 and 0.2 is used for DNN-1 and DNN-2, respectively.}
	\vspace{-0.2cm}
	\begin{center}
		%\footnotesize
		\begin{tabular}{|>{\arraybackslash}m{0.35in}|>{\centering\arraybackslash}m{0.25in}|>{\centering\arraybackslash}m{0.55in}|>{\centering\arraybackslash}m{0.5in}|>{\centering\arraybackslash}m{0.01in} @{}m{0pt}@{}}
			\cline{1-4}
			{Layer} & {Size} & {Activation} & {Dropout} & \parbox{0pt}{\rule{0pt}{2ex+\baselineskip}} \\ [0ex]  	% [1ex] adds vertical spac
			\cline{1-4}
			fc1 & 512 & ReLU & 0.5/0.2 & \parbox{0pt}{\rule{0pt}{0ex+\baselineskip}} \\ [0ex]  	% [1ex] adds vertical spac
			
			fc2 & 256 & ReLU & 0.5/0.2 & \parbox{0pt}{\rule{0pt}{0ex+\baselineskip}} \\ [0ex]  	% [1ex] adds vertical spac
			
			fc3 & 512 & ReLU & 0.5/0.2 & \parbox{0pt}{\rule{0pt}{0ex+\baselineskip}} \\ [0ex]  	% [1ex] adds vertical spac
			\cline{1-4}
		\end{tabular}
	\end{center}
	\label{tab:dnn}
	\vspace{-0.5cm}
\end{table}
%-----------------------------------

\vspace{-0.2cm}
\subsection{DNN-2: Joint event classification and boundary estimation with multitask loss}
\label{ssec:event_classification}
\vspace{-0.1cm}
We enforce DNN-2 to jointly model the class distribution for event classification and the event temporal structures for event boundary estimation, similarly to \cite{Phan2015,Phan2016e}. 
The proposed multi-task loss is specialized for this purpose. Multitask modelling can also be interpreted as implicit regularization, which is expected to improve generalization of a network \cite{Redmon2016,Ruder2017}. 

In addition to the one-hot encoding vector $\mathbf{y} \in \{0,1\}^C$, where $C$ is the number of event categories, we associate a sample $\mathbf{x}$ with a distance vector $\mathbf{d}~=~(d_{\mbox{\scriptsize on}}, d_{\mbox{\scriptsize off}}) \in \mathbb{R}^2$. $d_{\mbox{\scriptsize on}}$ and $d_{\mbox{\scriptsize off}}$ denote the distances from the audio frame $\mathbf{x}$ to the corresponding event onset and offset \cite{Phan2015,phan2015b}. The distances are further normalized to $[0,1]$ by dividing by their maximum values in the training data.

\begin{figure} [!t]
	\centering
	\includegraphics[width=0.7\linewidth]{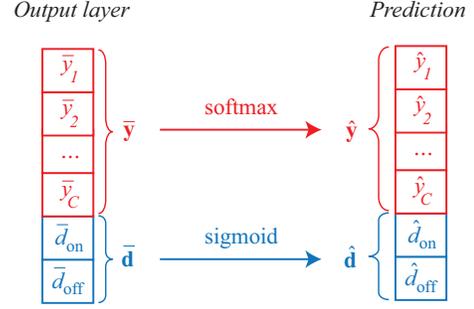}
	\vspace{-0.2cm}
	\caption{The output layer and the prediction of the multi-task DNN.}
	\label{fig:output_layer_multitask}
	\vspace{-0.3cm}
\end{figure}

The output layer of DNN-2 consists of two variables: $\bar{\mathbf{y}}=(\bar{y}_1, \bar{y}_2, \ldots, \bar{y}_C)$ and $\bar{\mathbf{d}}=(\bar{d}_{\mbox{\scriptsize on}}, \bar{d}_{\mbox{\scriptsize off}})$ as illustrated in Fig. \ref{fig:output_layer_multitask}. The network predictions for class posterior probability $\hat{\mathbf{y}}=(\hat{y}_1, \hat{y}_2, \ldots, \hat{y}_C)$ and distance vector $\hat{\mathbf{d}}=(\hat{d}_{\mbox{\scriptsize on}}, \hat{d}_{\mbox{\scriptsize off}})$ are then obtained by:
%\vspace{-0.1cm}
\begin{align}
\hat{\mathbf{y}} &= \softmax(\bar{\mathbf{y}}), \\
\hat{\mathbf{d}} &= \sigmoid(\bar{\mathbf{d}}).
\vspace{-0.05cm}
\end{align}

Given a training set $\left\{\left(\mathbf{x}_1,\mathbf{y}_1,\mathbf{d}_1\right),\ldots,\left(\mathbf{x}_N,\mathbf{y}_N, \mathbf{d}_N\right)\right\}$ of $N$ examples, DNN-2 is trained to minimize the multi-task loss:
\vspace{-0.05cm}
%\begin{align}
%E_{\mbox{\scriptsize mt}}(\boldsymbol{\theta}) =~&\lambda_{\mbox{\scriptsize class}}E_{\mbox{\scriptsize class}}(\boldsymbol{\theta}) + \lambda_{\mbox{\scriptsize dist}}E_{\mbox{\scriptsize dist}}(\boldsymbol{\theta}) \nonumber \\ 
%&+ \lambda_{\mbox{\scriptsize conf}}E_{\mbox{\scriptsize conf}}(\boldsymbol{\theta}) + \frac{\lambda}{2}\left\|\boldsymbol{\theta}\right\|^2_2,
%\label{eq:multitask_loss}
%\end{align}
\begin{align}
	E_{\mbox{\scriptsize mt}}(\boldsymbol{\theta}) = \lambda_{\mbox{\scriptsize class}}E_{\mbox{\scriptsize class}}(\boldsymbol{\theta}) + \lambda_{\mbox{\scriptsize dist}}E_{\mbox{\scriptsize dist}}(\boldsymbol{\theta}) + \lambda_{\mbox{\scriptsize conf}}E_{\mbox{\scriptsize conf}}(\boldsymbol{\theta}) + \frac{\lambda}{2}\left\|\boldsymbol{\theta}\right\|^2_2,
	\label{eq:multitask_loss}
	\vspace{-0.6cm}
\end{align}
where
\vspace{-0.2cm}
\begin{align}
E_{\mbox{\scriptsize class}}(\boldsymbol{\theta}) &= -\frac{1}{N}\sum_{n=1}^{N}{\mathbf{y}_n\log\big(\mathbf{\hat{y}}_n(\mathbf{x}_n\,,\boldsymbol{\theta})\big)}, \label{eq:class_loss}\\
E_{\mbox{\scriptsize dist}}(\boldsymbol{\theta}) &= -\frac{1}{N}\sum_{n=1}^{N}\left\|\mathbf{d}-\hat{\mathbf{d}}_n\left(\mathbf{x}_n,\boldsymbol{\theta}\right)\right\|^2_2, \label{eq:dist_loss} \\
E_{\mbox{\scriptsize conf}}(\boldsymbol{\theta}) &= -\frac{1}{N}\sum_{n=1}^{N}\left\|\mathbf{y}_{n} - \hat{\mathbf{y}}_{n}\frac{I\left(\mathbf{d}_n,\hat{\mathbf{d}}_n\left(\mathbf{x}_n, \boldsymbol{\theta}\right)\right)}{U\left(\mathbf{d}_n,\hat{\mathbf{d}}_n\left(\mathbf{x}_n, \boldsymbol{\theta}\right)\right)} \right\|^2_2. \label{eq:conf_loss}
\end{align}
$E_{\mbox{\scriptsize class}}$, $E_{\mbox{\scriptsize class}}$, and $E_{\mbox{\scriptsize conf}}$ in the above equations are so-called \emph{class loss}, \emph{distance loss}, and \emph{confidence loss}, respectively. The terms $\lambda_{\mbox{\scriptsize class}}$, $\lambda_{\mbox{\scriptsize dist}}$, and $\lambda_{\mbox{\scriptsize conf}}$ represent the weighting coefficients for three corresponding loss types. The \emph{class loss} complies with the common cross-entropy loss to penalize classification errors whereas the \emph{distance loss} penalizes event onset and offset distance estimation errors. Furthermore, the \emph{confidence loss} penalizes both classification errors and distance estimation errors. The functions 
%$I\left(\mathbf{d},\hat{\mathbf{d}}\right)$ and $U\left(\mathbf{d},\hat{\mathbf{d}}\right)$ in (\ref{eq:conf_loss}) 
$I(\mathbf{d},\hat{\mathbf{d}})$ and $U(\mathbf{d},\hat{\mathbf{d}})$ in (\ref{eq:conf_loss}) calculate the intersection and the union of the ground-truth event boundary and the predicted one respectively, given by:
%\vspace{-0.1cm}
\begin{align}
I\left(\mathbf{d},\hat{\mathbf{d}}\right) &= \min\left(d_{\mbox{\scriptsize on}}, \hat{d}_{\mbox{\scriptsize on}}\right) + \min\left(d_{\mbox{\scriptsize off}}, \hat{d}_{\mbox{\scriptsize off}}\right), \\
U\left(\mathbf{d},\hat{\mathbf{d}}\right) &= \max\left(d_{\mbox{\scriptsize on}}, \hat{d}_{\mbox{\scriptsize on}}\right) + \max\left(d_{\mbox{\scriptsize off}}, \hat{d}_{\mbox{\scriptsize off}}\right).
\end{align}

%\vspace{-0.3cm}
While the network may favour optimizing the class loss or the distance loss to reduce the total loss $E_{\mbox{\scriptsize mt}}(\boldsymbol{\theta})$, the confidence loss encourages it to optimize both losses at the same time. This is expected to accelerate and facilitate the learning process.

\vspace{-0.2cm}
\subsection{Inference}
\label{ssec:Inference}
\vspace{-0.1cm}

%IVM - by replacing DNN-1 and DNN-2 with the first DNN and the second DNN, I have increased the visual difference between the previous paper and the current paper...
Our prior work~\cite{Phan2017e} employed a simple inference step for event detection, which only used the output of one DNN plus the predicted class labels of a second DNN, combined with median filtering for label smoothing. The estimated event boundaries were completely ignored. Furthermore, that inference scheme could not guarantee reliably early event detection ability since the detection function did not fulfil the monotonicity requirement~\cite{Hoai2014,phan2015b} .
%We opted for a simple inference scheme here for target event segmentation. Firstly, we performed thresholding on the posterior probability output by the background-rejection classifier. A sample classified with a foreground posterior probability above a threshold $\alpha_{\mbox{\scriptsize prob}}$ will be subsequently forwarded to the event classifier to determine the event class label. Afterwards, an output label sequence was then smoothed by a median filter with a window length $w_{\mbox{\scriptsize sm}}$.
In this paper we propose an inference scheme that utilizes all available predicted quantities about target events. We will prove the monotonicity of the detection function, which is key to enabling the network to reliably detect target events early in time.

%Note that we did not use the estimates for event onset and offset distances provided by the event classification network. This can be further explored in future work as in \cite{Phan2015,phan2015b}.

\begin{figure} [!t]
	\centering
	\includegraphics[width=0.95\linewidth]{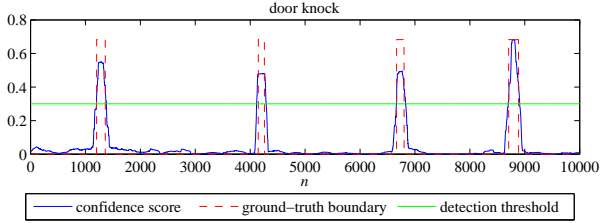}
	\vspace{-0.3cm}
	\caption{Alignment of the ``door knock'' confidence score and the ground-truth boundaries. The regions of confidence score above the threshold are considered as detected ``door knock'' events.}
	\label{fig:score_vs_threshold}
	\vspace{-0.3cm}
\end{figure}

Let $n,m>0$ both denote the frame time indices. Given a test audio frame $\mathbf{x}_m$ at the time index $m$, the time index $n$ is considered to be in the \emph{region of interest} (ROI) of the network prediction if the following condition is fulfilled:
\begin{align}
\mbox{~~~} m-\hat{d}_{\mbox{\scriptsize on}}(\mathbf{x}_m) \leq n \leq m+\hat{d}_{\mbox{\scriptsize off}}(\mathbf{x}_m),
\label{eq:region_of_interest}
\end{align}
where $\hat{d}_{\mbox{\scriptsize off}}(\mathbf{x}_m)$ and $\hat{d}_{\mbox{\scriptsize off}}(\mathbf{x}_m)$ represent the event onset and offset distances predicted by DNN-2. Note that these predicted distances need to be restored their original scales beforehand. The confidence score that a target event of class $c \in \{1, 2,\ldots, C\}$ occurs at the time index $n$ is then given by,
\begin{align}
f_c(n\,|\,\mathbf{x}_m) = \left\{ \begin{array}{lcl}
P_{1}\left(1\,|\,\mathbf{x}_m\right)P_{2}\left(c\,|\,\mathbf{x}_m\right) & \mbox{if} & (\ref{eq:region_of_interest}) \mbox{~holds}, \\
0 & & \mbox{~otherwise~}.
\end{array}\right.
\label{eq:c3_confidence_score_one_frame}
\vspace{-0.2cm}
\end{align}
In (\ref{eq:c3_confidence_score_one_frame}), $P_{1}\left(1\,|\,\mathbf{x}_m\right)$ represents the posterior probability for $\mathbf{x}_m$ being classified as foreground by DNN-1 and $P_{2}\left(c\,|\,\mathbf{x}_m\right)$ denotes the posterior probability for $\mathbf{x}_m$ being classified as the target class $c$ by DNN-2. The confidence score obtained by the network predictions given all audio frames then reads as,
\begin{align}
f_c(n) = \sum_{m}f_c(n\,|\,\mathbf{x}_m).
\label{eq:c3_confidence_score}
\end{align}
Fig. \ref{fig:score_vs_threshold} demonstrates the confidence score obtained for ``door knock'' events occurring in a test audio signal of the ITC-Irst dataset \cite{Temko07}. A class-specific threshold $\beta_c$ is then applied to the confidence score $f_c(n)$ for detection purposes, as demonstrated in Fig. \ref{fig:score_vs_threshold}.

\vspace{-0.2cm}
\subsection{Monotonicity of the detection function}
\label{ssec:monotonicity}
\vspace{-0.1cm}

The monotonicity of the detection confidence score can be proved easily. Let $f_{\bar{m}}(n)$ denote the accumulated confidence score up to the current time index $\bar{m}>0$. That is, 
\begin{align}
f_{\bar{m}}(n) &= \sum_{m=1}^{\bar{m}}f(n\,|\,\mathbf{x}_{m}),
\label{eq:accumulated_confidence_score}
\end{align}
where $f(n)$ is given in (\ref{eq:c3_confidence_score}). Note that we ignore the class label here for simplicity. Formally, we then have,
\vspace{-0.2cm}
\begin{align}
	\nonumber
	f_{\bar{m}}(n) = \sum_{m=1}^{\bar{m}}f(n\,|\,\mathbf{x}_{m}) &\leq \sum_{m=1}^{\bar{m}}f(n\,|\,\mathbf{x}_{m}) + f(n\,|\,\mathbf{x}_{\bar{m}+1})
	\\ 
	&= \sum_{m=1}^{\bar{m}+1}f(n\,|\,\mathbf{x}_{m}) = f_{\bar{m}+1}(n) .
	\label{eq:monotonicity}
	\vspace{-0.1cm}
\end{align}

The monotonicity property is guaranteed since, from (\ref{eq:c3_confidence_score_one_frame}), $f(n\,|\,\mathbf{x}_{m})~\geq ~0$ for all $m \geq0$.
The monotonicity can be interpreted as: the more the detector knows about the target event, the higher confidence it gains about occurrence of the target event. As soon as the accumulated confidence score reaches a pre-determined detection threshold, the event is considered detected. It is unnecessary for the system to see the entire event before triggering the detection.

\begin{figure*} [!t]
	\centering
	\includegraphics[width=0.875\linewidth]{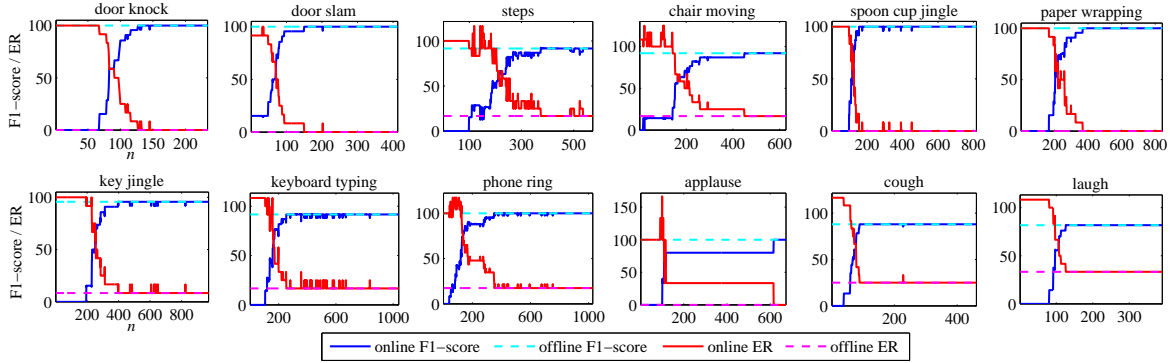}
	\vspace{-0.3cm}
	\caption{Development of the online performance curves of different categories. The offline performances are used as reference.}
	\label{fig:online_performance}
	\vspace{-0.4cm}
\end{figure*}

\vspace{-0.3cm}
\section{Experiments}
\vspace{-0.2cm}

\subsection{Experimental setup}
\label{ssec:setup}
\vspace{-0.1cm}

{\bf Dataset.} We conducted experiments on the ITC-Irst dataset of the CLEAR 2006 challenge \cite{Temko07,Zieger2005}. The data was recorded in a meeting-room environment with twelve recording sessions in total. There are 16 semantic event categories each of which has approximately 50 occurrences in the recordings. Following the CLEAR 2006 challenge setup \cite{Temko07}, twelve out of 16 event categories were evaluated while the rest was considered as background. Furthermore, nine out of twelve recordings were used for training and the remaining three were used for evaluation. Only the single channel named \emph{TABLE\_1} was used in the experiments.
%This dataset was also used in out previous work in early event detection \cite{phan2015b}. 

{\bf Features.} An audio signal was decomposed into frames of length 100 ms with a hop size of 10 ms. 64 log-Gammatone spectral coefficients \cite{Ellis2009} in the frequency range of 50 Hz to 22050 Hz were then extracted for each frame. In addition, we considered a context of five frames for classification purpose. The feature vector
for a context window was formed by simply concatenating feature vectors of its five constituent frames.

{\bf Parameters.} For the weighted loss in (\ref{eq:weighted_loss}), we set $\lambda_{\mbox{\scriptsize fg}} = 2$ and $\lambda_{\mbox{\scriptsize bg}} = 1$ which are inversely proportional to the ratio of foreground and background examples in the training data. 
As a result, false negatives are penalized twice as much as false positives. The associated weights of the multi-task loss in (\ref{eq:multitask_loss}) were set to $\lambda_{\mbox{\scriptsize class}} = 1$, $\lambda_{\mbox{\scriptsize dist}}= 2$, and $\lambda_{\mbox{\scriptsize conf}} = 1$. We set $\lambda_{\mbox{\scriptsize dist}}$ larger than $\lambda_{\mbox{\scriptsize class}}$ and $\lambda_{\mbox{\scriptsize conf}}$ to encourage DNN-2 to focus more on modelling event temporal structures. In addition, we set the regularization parameter $\lambda = 10^{-3}$ for both losses. The DNNs were trained using the \emph{Adam} optimizer \cite{Kingma2015} with a learning rate of $10^{-4}$. DNN-1 was trained for 25 epochs with a batch size of 256 whereas DNN-2 was trained for 25 epochs with a batch size of 128.

The detection confidence scores were normalized to $[0,1]$ and the class-wise detection thresholds were searched in the range of $[0,1]$ with a step size of $0.1$ via 9-fold cross-validation on the training data. Those threshold values which yielded maximum class-specific F1-scores were retained. 

{\bf Baseline.} For comparison, we used the early event detection system based on random regression forests proposed in \cite{phan2015b} as the baseline. The setting for the random-forest classification and random regression forests was unchanged from~\cite{phan2015b} apart for a minor improvement in the inference step. That is, we allowed audio frames to contribute to boundary estimations of all event classes instead of a single class as in \cite{phan2015b}. The class-wise contribution was weighted by the posterior probability that the frame is classified into the corresponding class. The baseline was run for five times and its average performance was reported.

{\bf Evaluation metrics.} Two metrics were used for evaluation: event-wise detection error rate (ER) and event-wise detection accuracy in terms of F1-score.

%Furthermore, to illustrate the efficiency of our proposed approach, we also compare the obtained performance to those achieved by the multi-channel system in \cite{phan2015d} and the verified detection system in \cite{phan2017d}

\begin{table} [t]
	\vspace{-0.15cm}
	\caption{ER (\%) and F1-score (\%) obtained by different detection systems. Bold denotes where the proposed system performed equally to or better than the baseline.}
	%\small
	\vspace{-0.2cm}
	\begin{center}
		%\begin{tabular}{|c|c|c|c|c|}
		\begin{tabular}{|>{\arraybackslash}m{0.9in}|>{\centering\arraybackslash}m{0.4in}|>{\centering\arraybackslash}m{0.4in}|>{\centering\arraybackslash}m{0.4in}|>{\centering\arraybackslash}m{0.4in}|}
			\hline
			\multirow{2}{*}{\bf Event Type}  & \multicolumn{2}{c|}{\bf ER} & \multicolumn{2}{c|}{\bf F1-score} \parbox{0pt}{\rule{0pt}{1ex+\baselineskip}}\\ 	% inserts table heading
			\cline{2-5}
			&  Reg. Forests & DNN &  Reg. Forests & DNN \parbox{0pt}{\rule{0pt}{0ex+\baselineskip}}\\ 	% inserts table heading
			\hline
			door knock &  $1.7$  & $\mathbf{0.0}$ & $99.2$ & $\mathbf{100.0}$ \parbox{0pt}{\rule{0pt}{0ex+\baselineskip}} \\ 	% [1ex] adds vertical spac
			%\hline
			door slam & $0.0$ & $\mathbf{0.0}$ & $100.0$ & $\mathbf{100.0}$ \parbox{0pt}{\rule{0pt}{0ex+\baselineskip}}\\ 	% [1ex] adds vertical spac
			%\hline
			steps & $15.0$ & $16.7$ & $92.5$ & $91.7$ \parbox{0pt}{\rule{0pt}{0ex+\baselineskip}}\\ 	% [1ex] adds vertical spac
			%\hline
			chair moving &  $48.3$ & $\mathbf{16.7}$ & $84.4$ & $\mathbf{92.0}$ \parbox{0pt}{\rule{0pt}{0ex+\baselineskip}}\\   	% [1ex] adds vertical spac
			%\hline
			spoon cup jingle & $3.3$ & $\mathbf{0.0}$ & $98.3$ & $\mathbf{100.0}$ \parbox{0pt}{\rule{0pt}{0ex+\baselineskip}}\\   	% [1ex] adds vertical spac
			%\hline
			paper wrapping  & $0.0$ & $\mathbf{0.0}$ & $100.0$ & $\mathbf{100.0}$ \parbox{0pt}{\rule{0pt}{0ex+\baselineskip}}\\ 	% [1ex] adds vertical spac
			%\hline
			key jingle & $5.0$ & $8.3$ & $97.7$ & $95.7$ \parbox{0pt}{\rule{0pt}{0ex+\baselineskip}}\\  	% [1ex] adds vertical spac
			%\hline
			keyboard typing & $41.7$ & $\mathbf{16.7}$ & $81.5$ & $\mathbf{91.7}$ \parbox{0pt}{\rule{0pt}{0ex+\baselineskip}}\\  	% [1ex] adds vertical spac
			%\hline
			phone ring &  $23.5$ & $\mathbf{17.4}$ & $89.9$ & $\mathbf{100.0}$ \parbox{0pt}{\rule{0pt}{0ex+\baselineskip}}\\	% [1ex] adds vertical spac
			%\hline
			applause & $0.0$ & $\mathbf{0.0}$ & $100.0$ & $\mathbf{100.0}$ \parbox{0pt}{\rule{0pt}{0ex+\baselineskip}}\\  	% [1ex] adds vertical spac
			%\hline
			cough & $16.7$ & $25.0$ & $92.2$ & $88.0$ \parbox{0pt}{\rule{0pt}{0ex+\baselineskip}}\\  	% [1ex] adds vertical spac
			%\hline
			laugh &  $16.7$ & $33.3$ & $90.9$ & $81.8$ \parbox{0pt}{\rule{0pt}{0ex+\baselineskip}}\\  	% [1ex] adds vertical spac
			\hline
			{\bf Overall} &  $15.1$ & $\mathbf{11.0}$ & $93.1$ & $\mathbf{95.2}$ \parbox{0pt}{\rule{0pt}{0ex+\baselineskip}} \\   	% [1ex] adds vertical spac
			\hline
		\end{tabular}
	\end{center}
	\label{tab:c3_offline_performance}
	\vspace{-0.6cm}
\end{table}

\vspace{-0.3cm}
\subsection{Experimental resutls}
\vspace{-0.1cm}

{\bf Offline performance.} The detection performance achieved by the proposed system as well as the baseline when entire events are seen by the systems are shown in Table \ref{tab:c3_offline_performance}. Overall, the proposed system significantly outperforms the regression-forest  baseline, improving the F1-score from $93.1\%$ to $95.2\%$ and reducing the ER from $15.1\%$ to $11.0\%$. Note that the performance of the regression-forest baseline reported here is slightly better than that in \cite{phan2015b} due to the improvement in the inference step as mentioned in Section \ref{ssec:setup}. The results obtained by the proposed system are even better than those obtained by the regression-forest system with verification (i.e. $94.6\%$ on F1-score \cite{phan2017d}).

In term of class-wise performance, the proposed DNN system performed reasonably well on most of the target event categories even though the employed DNN architecture is relatively simple. 
Interestingly, events that involve the human vocal tract were the exception, which we postulate is due to the lack of feature invariance caused by vocal tract length variation. A more complex neural network, e.g. a CNN~\cite{Phan2017e}, would likely circumvent this issue. However, fine-tuning the network architecture is not the main focus of this work, it is primarily the early detection capability. %Obviously, other types of networks, such as CNN or RNN, can be further explored. 
 %except for ``cough'' and ``laugh''. The obtained results on these human-generated events are significantly worse that those obtained by the baseline. Perhaps, due to the lack of feature invariance, the DNNs are inefficient to handle the vocal-tract length variation between actors who produced these events. Alternatively, using CNNs in replacement for the DNNs would be potential to overcome this shortcoming as shown in \cite{Phan2017e}. However, finding a network architecture which works best on this dataset is not the main focus of this work.

{\bf Online performance.} To verify the early detection ability of the proposed system, as in \cite{phan2015b}, a test audio stream was simulated as a sequence of audio frames coming to the system sequentially one-by-one. As a new event frame was available, the detection performance was re-evaluated and recorded. The offline performance was used as reference. Fig. \ref{fig:online_performance} illustrates how the online performance curves develop as functions of the number of observed event frames. 
As expected, for all categories, as more event frames are seen by the system, the online F1-scores continually increase while the online ERs scores decrease, until both match the offline scores. 
More importantly, the online curves always reach the offline ones before the events end, meaning that those events detectable by the system are always detected before they finish. Consider the ``laugh'' category as an example; about $50\%$ of events are correctly detected within the first 100 audio frames (equivalent to 1.0 seconds). 
The curve reaches the offline F1-score (i.e. 81.8\%) after observing about 120 frames (equivalent to 1.2 seconds). As ``laugh'' events last for approximately 400 frames, the online system needs less than 30\% of the event intervals to achieve the same detection accuracy as the offline system.

{\bf Discussion.} One can see small fluctuations of the online F1-score and ER curves from Fig. \ref{fig:online_performance}. This does not mean that confidence scores themselves fluctuate. In fact, this highlights the way an audio event is considered to be detected: the center of the detected event must fall inside the corresponding ground-truth event and vice versa. Furthermore, with the proposed inference scheme, the region of segmented events is quite narrow, particularly at their early stages. As a result, they can be off-center with respect to the ground-truths and hence counted as detection errors. This is valid for applications in which event segmentation is unimportant, but does lead to room for further investigation into better segmentation strategies.
%\textcolor{red}{As a result, during the course, this condition does not meet for some events as their centers shift}. However, it is still useful for applications in which correct event segmentation is not important. This open room for further investigation for a better segmentation strategy.

\vspace{-0.4cm}
\section{Conclusions}
\label{sec:refs}
\vspace{-0.3cm}

We presented an AED system which is able to infer ongoing events in audio streams and reliably detect them at their early stages. The key components of the proposed system are a pair of tailored-loss DNNs coupled with a novel inference scheme. The weighted-loss DNN was designed to cope with unbalanced foreground/background classification while the multitask-loss DNN was encouraged to jointly model event class distribution and event boundary estimation. Finally, the inference step made use of the network outputs to compute a confidence score that a target event is occurring at a certain time index. The monotonicity of the detection function, needed for reliable early detection, was also proved. Experiments on the standard ITC-Irst dataset yielded not only state-of-the-art detection performance, but also a demonstration of reliable early detection abilities of the proposed system.

\balance

% References should be produced using the bibtex program from suitable
% BiBTeX files (here: strings, refs, manuals). The IEEEbib.bst bibliography
% style file from IEEE produces unsorted bibliography list.
% -------------------------------------------------------------------------
\bibliographystyle{IEEEbib}
\bibliography{reference}
\end{document}